\documentclass[10pt,preprint,flushrt]{aastex}
\input psfig.sty

\newcommand{\kms}          {\mbox{${\rm km~s^{-1}}$}}
\newcommand{\cc}           {\mbox{${\rm cm^{-3}}$}}
\newcommand{\e}            {\mbox{$^{-1}$}}
\newcommand{\ee}           {\mbox{$^{-2}$}}

\newcommand{\simgt}        {\gtrsim}
\newcommand{\simlt}        {\lesssim}
\def\cm2{\mbox{${\rm cm^{-2}}$}}
\def\h2{\mbox{${\rm H}_2$}}
\def\nh2{\mbox{$n_{\rm H_2}$}}
\def\Nh2{\mbox{$N_{{\rm H}_2}$}}
\def\Mh2{\mbox{$M_{{\rm H}_2}$}}
\def\n2hp{\mbox{N$_2$H$^+$}}
\def\hcop{\mbox{HCO$^+$}}
\def\h13cop{\mbox{H$^{13}$CO$^+$}}

\begin{document}

\title{Gas flows around two young stellar clusters in NGC2264}
\author{Jonathan P. Williams and Catherine A. Garland}
\affil{Astronomy Department, University of Florida, Gainesville, FL 32611}
\email{williams, garland@astro.ufl.edu}

\shorttitle{Gas flows in NGC2264}
\shortauthors{Williams \& Garland}

\begin{abstract}
Observations of the dust and gas toward two young stellar clusters,
IRS1 and IRS2, in the NGC2264 star forming region are presented.
Continuum emission is used to locate the dusty envelopes around
the clusters and individual protostars within
and line emission from the $J=3-2$ transitions
of \hcop\ and \h13cop\ is used to diagnose the gas flows around them.
The molecular abundance, velocity centroid and dispersion are
approximately constant across the IRS1 clump. With these constraints,
the self-absorbed \hcop\ lines are modeled as a large scale collapse,
with speed $v_{\rm in}=0.3$~\kms\ and mass infall rate
$\dot M=4\times 10^{-4}~M_\odot$~yr\e, falling onto an expanding central
core. The signature of large scale collapse, with a similar speed
and mass infall rate, is also found toward IRS2 but again appears
disrupted at small scales. Individual protostars are resolved in this
cluster and their size and velocity dispersion show that the stellar system
is currently bound and no older than $5\times 10^5$~yr, but is destined
to become unbound and disperse as the surrounding cloud material is lost.
\end{abstract}


\section{Introduction}
The mass of the molecular ISM is dominated by the largest clouds which
tend to form stars in groups, rather than in isolation \citep{lada94}.
Star formation is inefficient, however, and as the binding mass of a
molecular cloud is lost through the action of stellar winds, radiation,
and outflows, most groups evaporate \citep{adams00} and become isolated
stellar systems.
Thus an increased understanding of clustered star formation is essential
for learning about the origin of most Galactic disk stars, perhaps
including the Sun.

In this paper, we present millimeter wavelength observations of
two star formation sites, IRS1 and IRS2, in the NGC2264 region of
northern Monoceros. Lying at a distance of 760~pc \citep{sung97},
this region is associated with a moderate mass cloud,
$M\simeq 3\times 10^4~M_\odot$ \citep{oliver96}, that contains 30 IRAS
sources, most classified as Class I protostars \citep{margulis89},
and $\sim 360$ near-infrared sources \citep{lada93}.

IRS1, also known as Allen's source \citep{allen72}, is an early B star
situated just north of the Cone nebula. \citet{thompson98} postulate that
it has triggered the formation of several low mass stars around it and
\citet{ward00} map a number of nearby massive dust condensations suggesting
that high mass star formation may also be occuring. Molecular spectral
lines show signatures of both outflow \citep{margulis88,schreyer97}
and infall \citep{wolf97,williams99a}.

IRS2, discovered by \citet{sargent84}, is a less luminous and less well
studied source $\sim 6'$ to the north of IRS1. A small cluster and
nebulosity is visible in the I band image of \citet{lada93}.
\citet{margulis88} found a CO outflow toward this source,
\citet{wolf95a} mapped an associated dense molecular clump
(which they called MonOB1-D),
and \citet{williams99a} found preliminary evidence for collapse.

Here, we combine continuum observations of the dusty envelopes around the
most embedded protostars with spectral line observations of the gas in
these two cluster forming regions. The molecular envelopes around the
two clusters have a similar mass but are morphologically distinct and
present an interesting comparative study. In both cases, the
self-absorbed \hcop\ line is used to diagnose relative motions
and show large scale collapse onto each stellar group that appears
to be disrupted by protostellar outflows on smaller scales.
In the bright, compact IRS1 clump, the optically thin \h13cop\ line is
used to map the molecular abundance, velocity centroid, and dispersion
and thereby constrain the modeling of the \hcop\ profiles.
In IRS2, where individual protostars are resolved, the size and
velocity dispersion is used to estimate an upper limit for their
formation time and to determine the gravitational stability of the group.
The observations are described in \S\ref{sec:obs}, and their analysis
in \S\ref{sec:analysis}. We conclude in \S\ref{sec:conclusion}.

\section{Observations}
\label{sec:obs}
The observations were carried out at the
Heinrich Hertz Telescope\footnotemark\footnotetext{the Heinrich Hertz
Telescope is operated by the Submillimeter Telescope Observatory on behalf
of Steward Observatory and the Max-Planck-Institut fuer Radioastronomie.}
on Mt. Graham, AZ over a four day period in April 2001.
We used the facility 19 channel bolometer to observe continuum emission
at $870~\mu$m and the 230~GHz receiver to observe the $J=3-2$ lines of
\hcop\ and \h13cop. The weather was good during the run and the zenith
opacity at $870~\mu$m ranged from 0.3 to 0.55;
for the spectral line observations system temperatures ranged from
350 to 450~K.

The bolometer observations were carried out in dual-beam raster
mode with an azimuth scanning velocity of $8''$/sec and a wobbler
frequency of 2~Hz.
Individual maps had sizes $\sim 8'\times 6'$, sampled at $10''$ in
azimuth and elevation, and took 40~minutes to complete.
Skydips were made before and after each map to determine the sky opacity
and observations of Jupiter and Uranus were used to calibrate the data.
Two maps were made of each source at different hour angles and therefore
different scan directions across the source. The data were reduced and
co-added using the NIC software package resulting in an rms noise in the
final maps of 0.14~Jy per $21''$ beam.

The spectral line data were also taken in on-the-fly mode.
Maps were made by scanning alternately in right ascension and declination
at $3''$/sec with the reference $12'$ west of the map center.
Individual map sizes were $\sim 3'\times 2'$, sampled at $10''$,
and took $\sim 25$~minutes each. System temperatures varied from
350~K to 425~K. Maps were repeated several times and
made at different offsets, then co-added and stitched together in the
CLASS software package so as to achieve an rms noise temperature of
0.8~K (\hcop) and 0.4~K (\h13cop) per 250~kHz ($\simeq 0.28$~\kms) channel.
The temperature scale was calibrated through the chopper wheel
method \citep{rohlfs96}, and we then applied main beam and
forward efficiencies $\eta_{\rm mb}=0.46,~\eta_{\rm fss}=0.92$.
Observations of the standard source Orion IRc2 were made daily as an overall
check of calibration, pointing, and setup. For both the bolometer and
line observations, pointing was checked through observations of Jupiter
every 2--3 hours.

\section{Results}
\label{sec:analysis}

\subsection{Bolometer data}
\begin{figure}[htpb]
\vskip -1.0in
\centerline{\psfig{figure=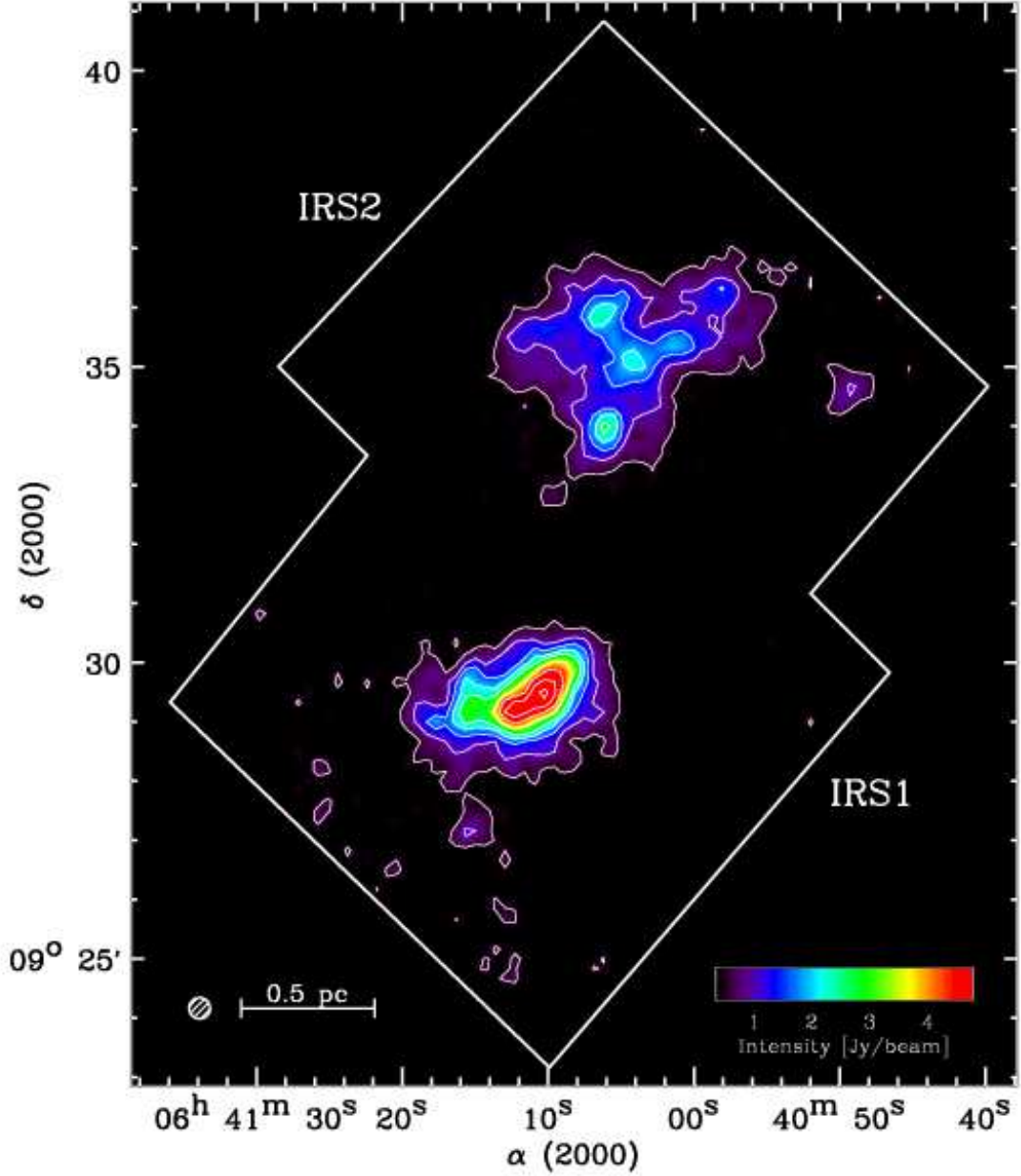,height=8.5in,angle=0,silent=1}}
\vskip -1.1in
\caption{$870~\mu$m continuum emission toward IRS1 and IRS2.
Black contours are at 0.5, 1.0, 1.5, 2.0, 2.5 Jy~beam\e,
and white contours are at 3.5, 4.5, 5.5, 6.5 Jy~beam\e.
The boundary of the map is marked by the heavy solid line and the
$21''$ beam size is indicated in the lower left corner.}
\label{fig:bolo}
\end{figure}

The $870~\mu$m continuum emission toward IRS1 and IRS2 is shown in
Figure~\ref{fig:bolo}. IRS1 is the bright, compact source to the
south. Its elongated shape shows signs of substructure;
\citet{ward00} resolve five distinct condensations with higher
resolution IRAM and JCMT observations. IRS2 lies to the north
and is more fragmented with several low intensity peaks, suggestive of
a slightly more evolved cluster in which the individual protostars are
beginning to shed their circumstellar envelopes and migrate away from
their birthplace.

The integrated fluxes, for flux densities greater than 0.5~Jy~beam\e,
of the two clusters are similar, 94~Jy for IRS1
(consistent with Ward-Thompson et al.) and 84~Jy for IRS2
indicating that they have similar dust masses and make a good
comparative study. The fluxes convert to masses through
$$M={F_\nu d^2\over \kappa B_\nu(T_d)}
   =8.4\left({F_\nu\over {\rm Jy}}\right)~M_\odot$$
where we have taken the mass opacity $\kappa=0.009$~cm$^2$~g\e,
dust temperature $T_d=17$~K, as in \citet{ward00}
and distance $d=760$~pc.
Thus the total interstellar mass around IRS1 is $790~M_\odot$
and $710~M_\odot$ around IRS2.
From the projected area of each group,
0.45~pc$^2$ for IRS1 and 0.62~pc$^2$ for IRS2,
we estimate average column and volume densities
$\langle\Nh2\rangle=8\times 10^{22}$~cm\ee,
$\langle\nh2\rangle=5\times 10^4$~\cc\ for IRS1 and
$\langle\Nh2\rangle=5\times 10^{22}$~cm\ee,
$\langle\nh2\rangle=3\times 10^4$~\cc\ for IRS2.

\subsection{Line data}
The average density of the material traced by the continuum
observations is similar to that of \hcop\ emitting gas.
Indeed, the morphology of the integrated line emission
follows the dust (see Figure~\ref{fig:irs1_h13cop})
and we therefore use the line data to investigate the
dynamics of the circumcluster material.

The optically thick \hcop\ spectra provide information on
relative motions, i.e. infall and outflow, in the gas
(e.g., \citet{walker86})
and the optically thin \h13cop\ show the intensity weighted
mean velocity and linewidth along each line of sight.
We also use the latter to measure the molecular abundance
in the IRS1 clump.

The individual protostars in the compact IRS1 clump merge
together in the $29''$ beam of the line observations and we
examine average, group properties. Objects in the more
extended IRS2 clump are resolved, however, and we can
determine mass infall rates onto both the cluster and
a single protostar. In addition, by comparing the cluster's
size and velocity dispersion, we estimate its age.

\subsubsection{IRS1}
\label{sec:irs1}
\begin{figure}[htpb]
\vskip -0.9in
\centerline{\psfig{figure=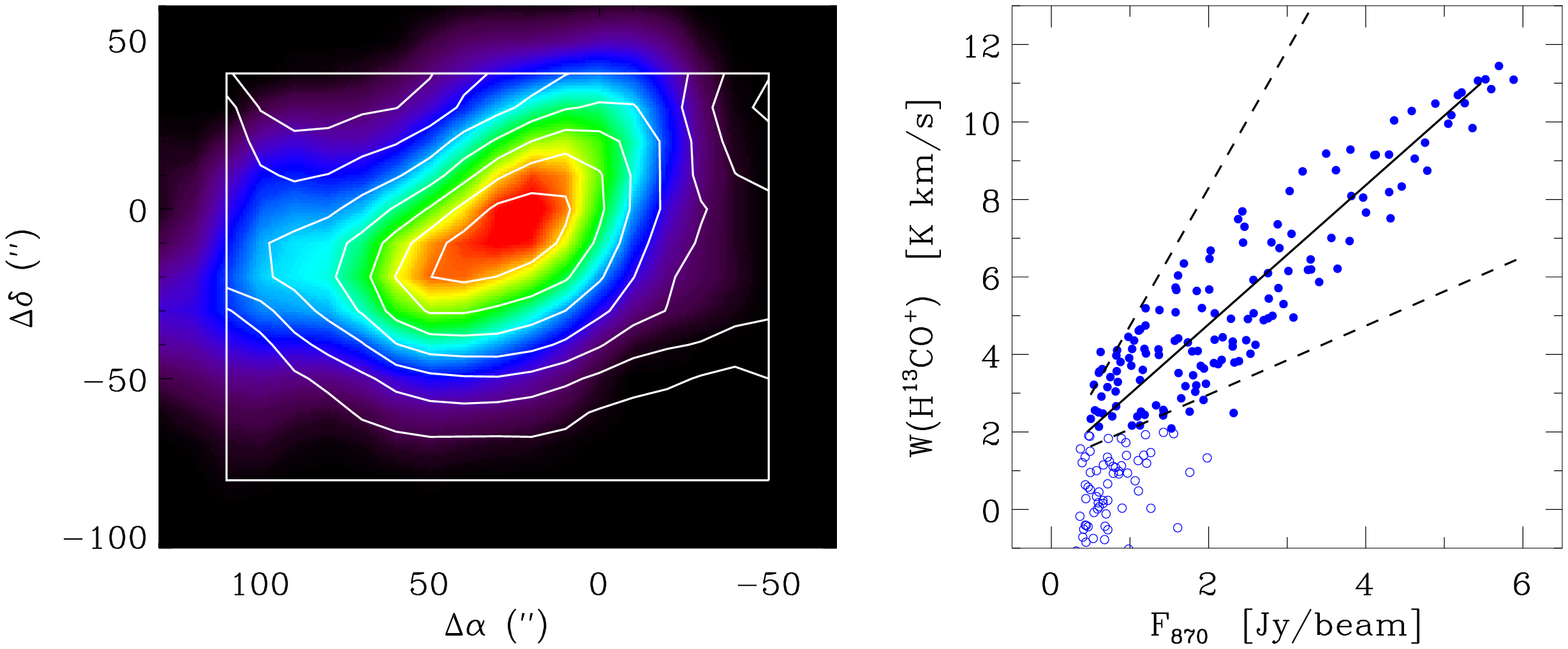,height=5.3in,angle=90,silent=1}}
\vskip -1.8in
\caption{Comparison of line and continuum data in the IRS1 clump.
The left panel shows contours of \h13cop\ emission, integrated over
the range $v=6$ to 10~\kms, overlayed on a grayscale of the $870~\mu$
thermal continuum emission, smoothed to the $29''$ resolution of the
line data. Contours begin and are at intervals of 1.5 K~\kms,
and the grayscale ranges from 0.6 to 6.0 Jy~beam\e.
The close correspondence between the two is shown
on a point by point basis in the right panel which plots the integrated
\h13cop\ intensity versus $870~\mu$ emission. The linear correlation
indicates that the \h13cop\ emission is optically thin and its abundance
constant across the clump.
The solid line is a linear least squares fit to the filled circles,
$W(\h13cop)>2$~K~\kms, and implies an abundance $x(\h13cop)=2.0\times 10^{-11}$
relative to H$_2$. The dashed lines show abundances factors of 2 lower
and higher.}
\label{fig:irs1_h13cop}
\end{figure}

Line and continuum data in IRS1 are compared in Figure~\ref{fig:irs1_h13cop}.
The left panel overlays contours of integrated \h13cop\
intensity over a grayscale of the dust emission.
Their similar appearance demonstrates that we can use
the line data to examine motions of the circumcluster
material. The right panel plots the two quantities on
a point by point basis and shows an approximately linear
trend over a decade in scale with the implication that
the \h13cop\ is both optically thin and its abundance
constant across the clump. By converting each quantity
to a column density (assuming an excitation temperature
of 20~K for the gas and dust temperature of 17~K)
the best fit \h13cop\ abundance relative to H$_2$ is
found to be $x(\h13cop)=2.0\times 10^{-11}$.
For an isotopic ratio, $^{12}{\rm C}/^{13}{\rm C}=50-100$,
the implied \hcop\ abundance lies in the range,
$x(\hcop)=1-2\times 10^{-9}$,
similar to values in other OB star forming regions
\citep{bergin97,ungerechts97}.
The lack of variation across the clump is
demonstrated by the two dashed lines bracketing the
data at $x(\h13cop)=[1.0,4.0]\times 10^{-11}$.

\begin{figure}[htpb]
\vskip -1.5in
\centerline{\psfig{figure=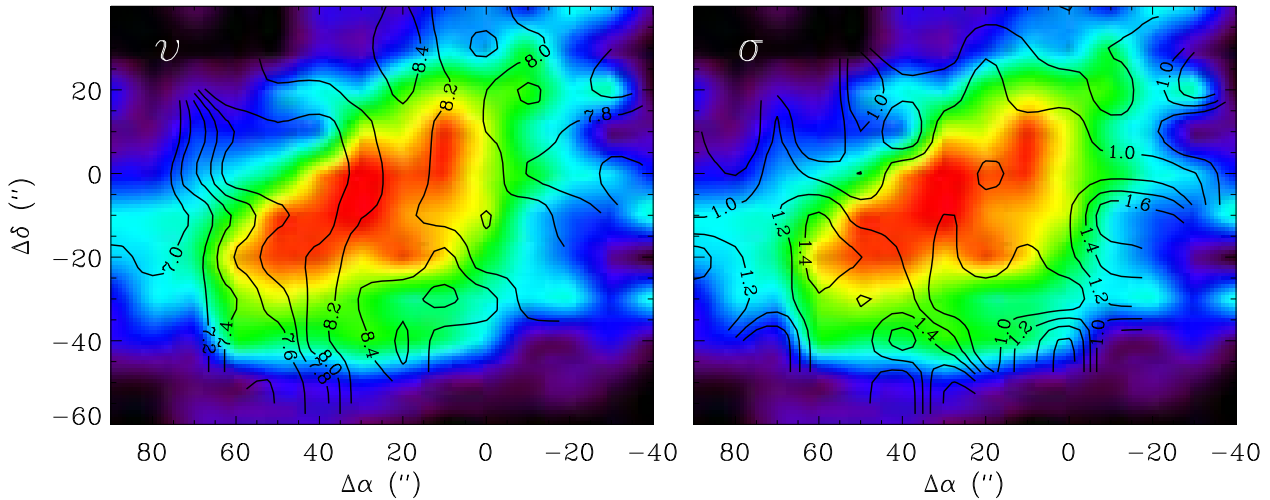,height=5.5in,angle=90,silent=1}}
\vskip -1.25in
\caption{Labeled contours of the mean velocity (left panel)
and velocity dispersion (right panel) of \h13cop\ in the IRS1 core,
overlayed on a grayscale of integrated \h13cop\ emission.
In the central region of the clump, $\Delta\alpha=-10''$ to $50''$,
the velocity and dispersion are both approximately constant.}
\label{fig:irs1_velocity}
\end{figure}

Higher moments of the \h13cop\ emission show the variation
of mean velocity and dispersion across the cluster
(Figure~\ref{fig:irs1_velocity})
The former increases from 7.0~\kms\ toward the East
to 8.4~\kms\ at center, but then returns to a lower value,
$\sim 7.8$~\kms\ in the West.
There is no clearly defined rotation across the clump
and toward the central region, $\Delta\alpha=-10''$ to $50''$,
the velocity difference, $\sim 0.3~\kms$, is much less than
the dispersion. The dispersion map itself also shows little
variation in this central region and has an average value,
$\sigma=1.0$~\kms.

\begin{figure}[htpb]
\vskip -1.0in
\centerline{\psfig{figure=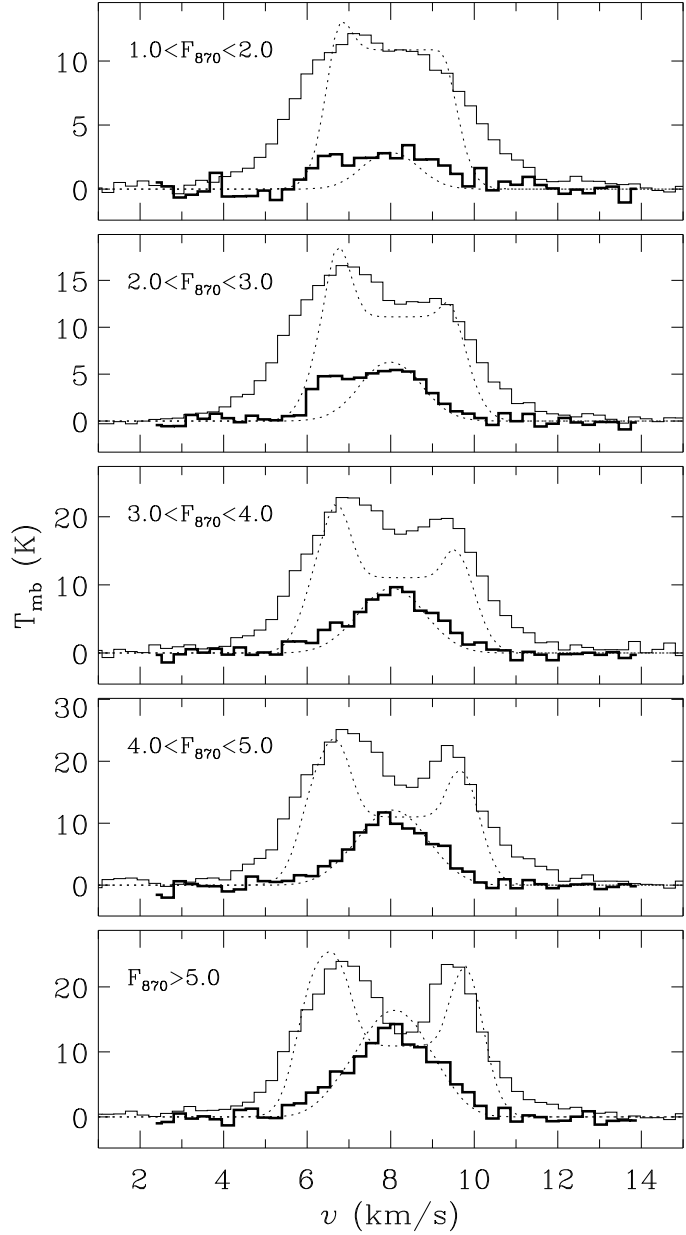,height=7.5in,angle=0,silent=1}}
\vskip -0.4in
\caption{\hcop\ and \h13cop\ 3--2 spectra toward IRS1 (light and heavy
lines respectively) averaged over different regions having common dust
continuum fluxes, $F_{870}$~Jy.
The \h13cop\ spectra are shown multiplied by three for clarity.
The upper panels are averaged over the outer parts of the
core with low continuum fluxes and show large scale infall. The lower panels
are averaged over the inner parts of the core with high continuum fluxes and
are more symmetric suggesting an overall balance between protostellar outflow
and core collapse. The dotted lines show model fits using a spherically
symmetric core, constrained by the dust temperature and column density
profile and \h13cop\ abundance, velocity, and linewidth, that consists
of an outer layer collapsing onto an expanding inner region.}
\label{fig:irs1_avespec}
\end{figure}

The \hcop\ spectra are self-absorbed across the clump: they
generally show two peaks and a central dip at the mean
velocity of \h13cop. Changes in the velocity of the absorbing
gas relative to the emitting gas alter the observed profile and
are a diagnostic of relative motions in the clump \citep{leung77}.
The relatively poor resolution
of the line data, $29''\simeq 0.1$~pc, does not permit a
detailed analysis of the dynamics around individual protostars
within this cluster and we therefore focus on global properties,
as illustrated in Figure~\ref{fig:irs1_avespec}.
Here, average spectra are shown, computed over different regions
of the clump corresponding to different $870~\mu$m flux densities.
This is similar to radial averaging but takes into account
the non-circular shape of the clump.
At low flux densities, corresponding to the outer regions,
the average spectra are relatively weak and the optical depths
are low. Nevertheless, the \hcop\ spectra are clearly asymmetric
and show the red-shifted self-absorption characteristic of
inward motions \citep{walker86}.
At higher flux densities, toward the clump center, the
line intensities and optical depths increase and the \hcop\
profiles show two peaks with a central dip. However, the
spectra become more symmetric with increasing flux density
until at the highest values at the clump's center, the two \hcop\
peaks are of similar intensity suggesting zero average inflow
in this region.

Infall spectra are notoriously difficult to interpret
\citep{menten87}, and the change in profile shapes observed
in IRS1 require detailed modeling to understand the flows of
material in this complex region. Since the clump has a single,
central peak and no clear velocity gradient, we use the 1-d
radiative transfer code, RATRAN, of \citet{hoger00}
to model, simultaneously, the dust and line data.
The continuum and \h13cop\ observations fix the
size, temperature, density profile, \h13cop\ abundance, mean
velocity, and velocity dispersion and tightly constrain the
model, leaving only the velocity profile and isotopic ratio
as the remaining parameters in the model output \hcop\ line profiles.

The model consists of 10 logarithmically spaced, radially symmetric,
shells from $r=10^{16}$~cm to $r=r_{\rm max}=10^{18}$~cm (0.32~pc).
\citet{ward00} show that the spectral energy distribution of
IRS1 is well fit by the sum of two greybodies consisting of a
compact component at 38~K and a larger, cooler component at
17~K that dominates the mass. Based on this, we set the temperature
of the outer layer at 17~K and the inner layers, $r<6\times 10^{17}$~cm
at 38~K. The gas temperature is fixed to be the same as the dust
as would be expected at the high densities in the model.
Figure~\ref{fig:irs1_h13cop} shows that the clump is moderately,
but not strongly, peaked and we find that the density profile
$\nh2(r)=10^5(r_{\rm max}/r)^{3/2}$~\cc, when convolved with a $29''$
beam, matches the dust and velocity integrated \h13cop\ maps well.
The abundance, $x(\h13cop)=2.0\times 10^{-11}$,
systemic velocity, $v=8.0$~\kms, and velocity dispersion,
$\sigma=0.8$~\kms, are set constant in all layers as indicated
by the \h13cop\ data. The high relative motions between the
layers contribute to broaden the output \h13cop\ spectra close
to the observed 1.0~\kms. With these parameters, we find that
we can reproduce the (radially averaged) continuum fluxes
and \h13cop\ spectra to within 20\%.

Under these constraints, it is not possible to fit the observed
\hcop\ spectra as closely since only the isotopic abundance and
relative velocities between the layers can be varied.
Through experimentation with the models, it was found that,
given the observed density, temperature, abundance, and velocity
profiles, the increasing symmetry of the \hcop\ spectra toward
the cluster center can only be explained by an outer layer
collapsing onto an {\it expanding} inner region.
Figure~\ref{fig:irs1_avespec} plots radially averaged model
\hcop\ and \h13cop\ spectra as dotted lines in comparison with
the observed data for the case where the isotopic ratio,
$^{12}{\rm C}/^{13}{\rm C}=75$ (implying $x(\hcop)=1.5\times 10^{-9}$),
and the outer layer ($r=6\times 10^{17}$ to $1\times 10^{18}$~cm)
collapses at a speed $v_{\rm in}=0.3$~\kms\ onto the inner layers
($r=10^{16}$~cm to $6\times 10^{17}$~cm) expanding at a speed
$v_{\rm exp}=1.0$~\kms.

The combination of large scale collapse and central expansion
conspire to match the observed spectral variations from core edge
to center because the optical depth is very high in the low density,
outer layers and produces a broad (blue and redshifted),
self-absorption dip along lines of sight toward the center
but only redshifted self-absorption along lines of sight near
the edges that do not intercept the inner expanding layer.
The velocity of the absorption dip is not observed to change
from core center to edge and is reproduced in the model via the
increased projection of the radial flow along lines of sight
toward the edges.
The model spectra have broader self-absorption dips than the
data because of the small number of layers in the model and
the abrupt change in temperature from 17~K in the outer layer
to 38~K in inner layers. Applying a smoother temperature
transition and more layers produces more rounded self-absorption
dips but at the expense of adding additional parameters to the model.

Because of the radial dependence, $M(<r)\propto r^{3/2}$,
most of the mass of the core is in the outer layers and
altering the velocities of the innermost layers,
$r\simlt 10^{17}$~cm, does not change the output profiles
noticeably. That is, the model resembles the two layer model
described by \citet{myers96} but with a spherical geometry.
With that caveat, the output \hcop\ profiles are very
sensitive to the optical depth and infall speed of the outer layer
and expansion speed of the layer interior to that
(which together account for about three quarters of the total mass)
and through experimentation with the models we found that the
column density and velocities of these layers are
constrained to within $\sim 20\%$ by the requirement
of matching the observed redshifted absorption at large
clump radii and near-symmetry at small radii.

The outer, infalling, layer contains 1/2 the total mass,
or about $M_{\rm in}=400~M_\odot$, and the mass infall rate
onto the cluster is therefore
$$\dot M_{\rm in} = M_{\rm in}v_{\rm in}/r_{\rm max}
                  = 4\times 10^{-4}~M_\odot {\rm yr}^{-1}.$$
This is one to two orders of magnitude greater than infall rates
onto solar and intermediate mass protostars \citep{zhou95}
as expected for a global infall onto a group of such objects
and is similar to the value expected for a gravitational
flow, $\dot M=\sigma^3/G$, where $\sigma=1.0$~\kms\ is the average
velocity dispersion of the core (Figure~\ref{fig:irs1_velocity}),
and $G$ is the gravitational constant \citep{shu77}.

The model requires an expanding shell to fit the increasing
symmetry toward the core center. Physically, this is probably due
to the protostellar outflows in the region, one of which appears
to be directed along our line of sight \citep{schreyer97}.
\citet{shu87} hypothesize that (low mass) stars self-determine
their final mass through the removal of a collapsing envelope by
their outflows (see also \citet{velusamy98}).
By comparing clusters in different evolutionary states
and examining the gas flows around them, it will be possible to see
if such a process is relevant in a clustered environment where most
stars form, and therefore assess its role in the origin of the
stellar IMF \citep{adams96}.

\subsubsection{IRS2}
\begin{figure}[htpb]
\vskip -0.9in
\centerline{\psfig{figure=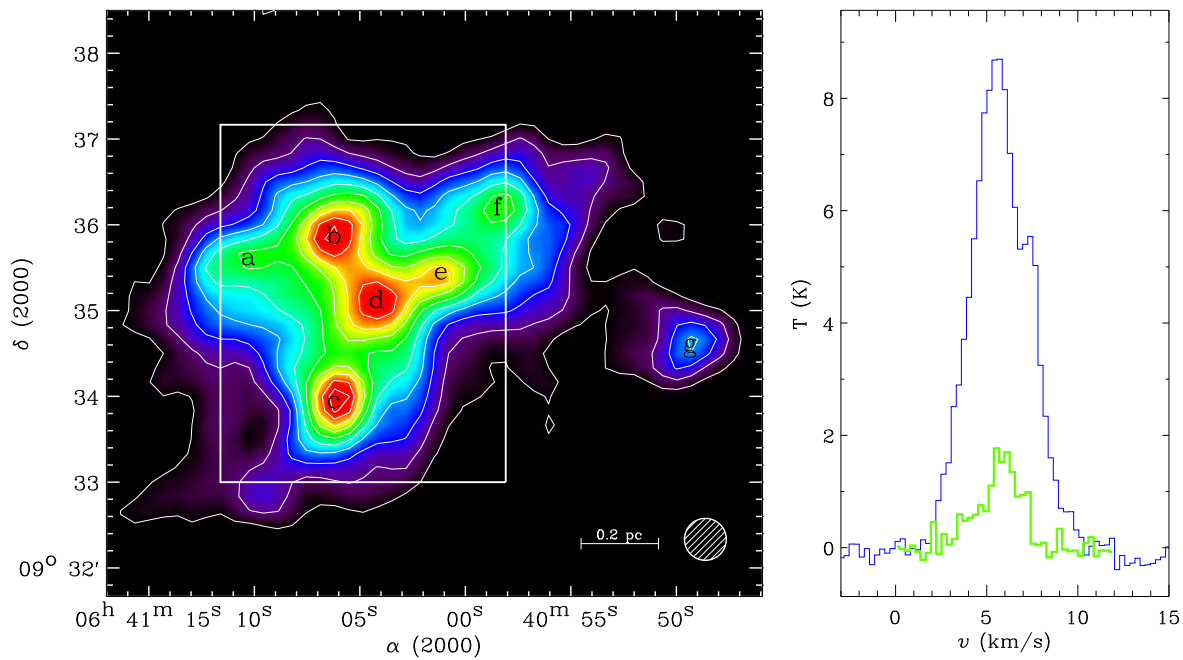,height=5.3in,angle=90,silent=1}}
\vskip -0.9in
\caption{Structure and dynamics in the IRS2 cluster. The left panel is a map
of the dust continuum emission smoothed to the spectral line resolution of 
$29''$ as indicated in the lower right corner. Contours begin at and are in
steps of 0.2 Jy~beam\e. Seven distinct peaks of emission are labeled a-g.
The solid dark line marks the boundary of the line data and the
white boxes within this outline the different regions over which the average
\hcop\ (light lines) and \h13cop\ ($\times 3$; heavy lines) spectra
in the three left panels, are computed.
As with IRS1, the spectra show that large scale inflow is arrested
on small scales around protostellar cores.}
\label{fig:irs2_avespec}
\end{figure}

Continuum and line emission were lower in IRS2 than in IRS1.
The proportion between them, however, was approximately the
same and the \h13cop\ abundance was found to be
consistent with that measured in IRS1, though with a greater
uncertainty due to the lower signal-to-noise ratio.
Due to the relative weakness of the emission, only the central
part of the cluster was mapped in \hcop\ and \h13cop.
Figure~\ref{fig:irs2_avespec} plots the boundary of the line
data on a map of continuum emission. Slight red-shifted
self-absorption is found in the average \hcop\ spectrum
shown in the left panel. The self-absorption dip is small
due to the low optical depth when averaged over the entire
mapped region, but is more pronounced in individual spectra,
particularly around the dust peaks where the column density
is greatest. Both red- and blue-shifted asymmetries are found
(see below) but the dominant tendency is for red-shifted
asymmetry suggesting, as in IRS1, large scale collapse onto
the cluster.

Given the complex resolved structure in IRS2, similar
fits to the spectra at different size-scales as in
\S\ref{sec:irs1}, are not feasible with a 1-d model.
However, both simple plane-parallel and spherical two layer
model fits to the average spectrum imply an infall speed
$v_{\rm in}\simeq 0.3\pm 0.1$~\kms.
This is similar to the speed of the collapsing layer
around IRS1 and implies a similar mass infall rate,
$\dot M_{\rm in} \simeq 4\times 10^{-4}~M_\odot {\rm yr}^{-1}$.

\begin{figure}[htpb]
\vskip -0.8in
\centerline{\psfig{figure=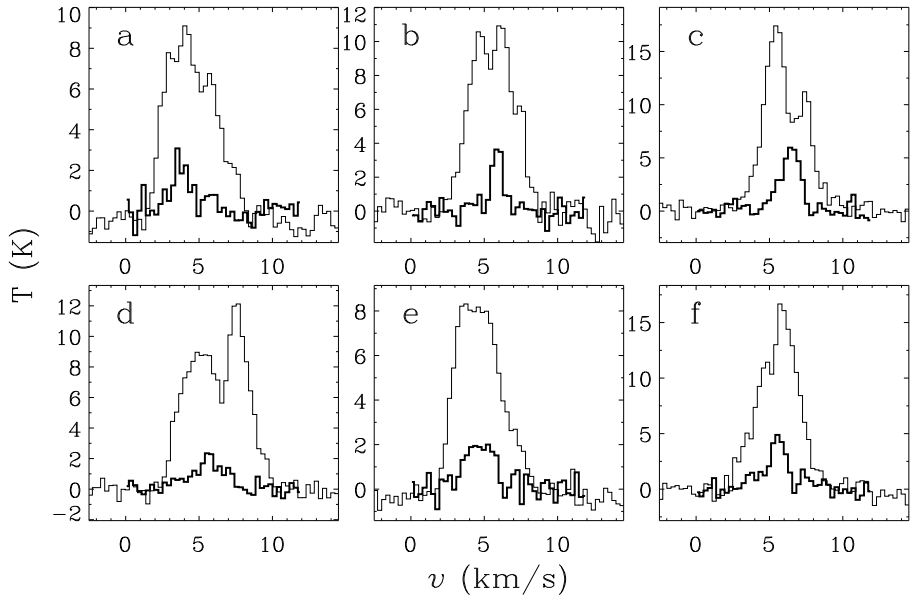,height=4.4in,angle=90,silent=1}}
\vskip -0.4in
\caption{\hcop\ (light lines) and \h13cop\ ($\times 3$; heavy lines) spectra
toward the individual IRS2 cores a-f as labeled in Figure~\ref{fig:irs2_avespec}.}
\label{fig:irs2_cores}
\end{figure}

The greater separation of the protostars in IRS2 allows an
investigation of the flows around each one individually.
Seven cores, labeled a-g in Figure~\ref{fig:irs2_avespec}
are apparent as peaks in the dust emission and interpolated
spectra (at a resolution of $29''$) toward the six that lie
within the boundary of the line data are shown in
Figure~\ref{fig:irs2_cores}. A complex mixture of dynamical
states are found. Core c has the classic infall profile
and core d shows a strong reversed signature. Cores e and f show
possible signs of infall and reversed flow respectively but the
optical depth is too low to be sure. The \hcop\ spectra toward cores
a and b are not readily interpreted in the same manner, due to
multiple peaks possibly from absorption by overlapping outflows.

Gaussian fits to the \h13cop\ spectra show that the non-thermal
velocity dispersion is typically in the range 0.7 to 1.0~\kms\
but has a pronounced local minimum of 0.4~\kms\ around core b.
The resolution of these data is too low to test collapse
models in detail but we do not find a correlation between the
\hcop\ asymmetry and the \h13cop\ dispersion, analogous to that
seen in the Serpens cluster \citep{williams00} and as might be
expected from motions driven by the dissipation of turbulence.

To estimate the dynamical influence of each protostar, we have
examined average spectra around each one with different smoothing
lengths. Although there is confusion from merging at angular
scales $\simgt 45''$, we find that the overall infall dynamics
of the cluster eventually dominate and that the influence of
any single protostar does not exceed a radius $60''\simeq 0.2$~pc.

Core c shows the clearest infall signature with a deep, red-shifted,
absorption dip. The infall spectra extend for $\sim 60''$ away from
the core before merging with the reversed profiles of spectra associated
with core d. The core also has the largest envelope with
a total continuum flux of 12.3~Jy and integrated \h13cop\ intensity
of 2.7~K~\kms\ within a radius $r=0.17$~pc, implying a mass
$M_{\rm env}=100~M_\odot$.
The high column density through this envelope is responsible for
the deep self-absorption dip, and does not imply a particularly
high infall speed.
A model fit using the spherically symmetric RATRAN code
implies an infall speed $v_{\rm in}=0.3$~\kms, and mass
infall rate $\dot M_{\rm in} = 2\times 10^{-5}~M_\odot {\rm yr}^{-1}$.
The speed is the same as the collapse around the entire cluster
but the mass infall rate is considerably less because the
fit requires only an outer $10~M_\odot$ layer to be collapsing.
Dividing the overall mass infall rate onto the cluster by the
number of protostars within it gives an average mass infall rate
per protostar about 3 times higher than measured around core c,
the one with the clearest individual collapse signature: clearly
not all the material falling onto the cluster makes it all the way
to the scale of protostars.

The one other core, d, with a deep self-absorption dip shows the
opposite asymmetry, indicative of an expanding, rather than infalling,
outer layer. It also has the largest \h13cop\ linewidth, 1.0~\kms,
in the region. Using a similar model as for core c, the expansion
speed is estimated to be 0.5~\kms. This is likely due to an
embedded star although there is no IRAS or MSX point source
at its center: higher sensitivity mid-infrared observations
are required to determine the stellar content of each core.

Because the envelopes around individual protostars in the cluster
are resolved, the optically thin \h13cop\ spectra give a measure
of the core-core velocity dispersion in the cluster.
The six cores, a-f, have a dispersion $\sigma_{\rm c-c}=0.90$~\kms\
and extend over an area with radius $r=0.44$~pc.
Together, these imply a virial mass,
$M_{\rm virial}=3r\sigma_{\rm c-c}^2/G=250~M_\odot$.
This is less than the total interstellar mass estimated
from the continuum observations but is likely much greater
than the final stellar mass of the system \citep{lada93}.
Thus, although the system is currently bound, as the surrounding
gas is dispersed through, e.g., stellar outflows, it is likely to
become unbound \citep{lada84}.

An unbound cluster expands and its age may be estimated from its
size and expansion rate. This effect is seen in optically visible
clusters in Orion over a range of ages $\sim 3-12$~Myr
\citep{blaauw91,brown97}
but the morphology of the two groups IRS1,2 in Figure~\ref{fig:bolo}
suggests that it may apply to considerably earlier times.
In IRS2, if $\sigma_{\rm c-c}$ is due to the expansion of the
protostellar group, the formation time can be estimated from
its crossing time, $t_{\rm cross}=r/\sigma_{\rm c-c}=5\times 10^5$~yr,
characteristic of late Class I sources \citep{mundy00}.
The gravitational attraction of the gas around the stellar group
should slow their expansion, and the crossing time is therefore
an upper limit to the cluster age.
Given the continued infall onto the cluster and the large
envelope mass, it is not clear if the cluster is actually expanding
and whether the crossing time is a useful measure of the cluster
age at these early stages in its formation but the suggestive
morphology and flux densities of the continuum maps are compelling.
Additional observations of other embedded stellar groups in this
and other regions will show if there is a consistent pattern.
In this regard, we note that a very short integration toward a
more extended (and therefore possibly more evolved) group of
IRAS sources $11'$ north of IRS2 did not detect $870~\mu$m
emission to an rms flux level of 0.1~Jy~beam\e.

\section{Conclusion}
\label{sec:conclusion}
We have mapped two embedded stellar groups, IRS1 and IRS2, in the
NGC2264 region in the continuum at $870~\mu$m and the $J=3-2$ lines
of \hcop\ and \h13cop. The former shows the thermal emission from
the dusty envelopes around the clusters and their protostellar members
and the latter show the flows around them; the optically thick
\hcop\ diagnosing relative motions (infall and outflow) and the
optically thin \h13cop\ revealing the bulk motion of the clusters
and the protostellar velocity dispersion within IRS2.

The two groups have a similar total mass, $M\simeq 10^3~M_\odot$,
but IRS1 has a much higher peak flux density and is more compact
than the more dispersed, lower surface brightness IRS2.
In IRS1, comparing the continuum flux density with the integrated
line intensity shows that the \h13cop\ abundance is constant across
the clump at the $29''$ resolution of these data.
Average \hcop\ spectra are fit with a radially symmetric model
constrained by the dust and \h13cop\ data and show a large scale
collapse with an infall speed $v_{\rm in}=0.3$~\kms\ and mass infall
rate $\dot M_{\rm in}=4\times 10^{-4}~M_\odot {\rm yr}^{-1}$,
falling onto an inner core expanding at speed $v_{\rm exp}=1.0$~\kms.
More detailed observations of the competing processes of
infall and outflow will show its role in determining stellar masses.

Large scale collapse with a similar infall speed and mass infall
rate are found in IRS2 but detailed modeling is precluded by the
complex structure that is resolved within. Flows around individual
protostars show a mixture of dynamical states ranging from localized
collapse with the same speed as the large scale collapse
and mass infall rate,
$\dot M_{\rm in}=2\times 10^{-5}~M_\odot {\rm yr}^{-1}$,
to expansion at speeds $v_{\rm out}\simeq 0.5$~\kms.
By averaging spectra over increasingly larger size scales,
the influence of any one protostar was found to be limited
to within a radius $\simlt 0.2$~pc.

The \h13cop\ spectra in IRS2 was used to measure the protostellar
velocity dispersion and thereby infer the virial mass of the
system. This was found to be less than the total mass enveloping
the cluster but more than the stellar mass indicating that the
group is destined to become unbound as the surrounding gas and
dust are dispersed by protostellar winds and outflows.
The expansion of an unbound stellar group may already have
started in IRS2 where the greater separation of individual
protostars and the low column density toward them,
relative to IRS1, both suggest a more evolved system.
In this case, an upper limit to its age, $\sim 5\times 10^5$~yr,
can be determined kinematically by assuming a constant expansion
speed equal to the observed protostellar velocity dispersion.
Additional observations of other clusters are necessary
to see if there is a pattern between cluster sizes, velocity
dispersion, and evolutionary state, but this technique
-- a well known tool for optical clusters --
may be a useful new approach for following protocluster and
protostellar evolution.

The dominant mode of stellar birth is in groups and these observations
provide new insight into their formation. To examine cluster formation
at earlier times will require higher resolution observations of dense,
compact cores such as IRS1, while large scale, high sensitivity imaging
of the dust and gas around more evolved systems than IRS2 can follow
the evolution at later times than discussed here.

\acknowledgments
We thank the scientific staff at the HHT for their assistance
during the observation run, Michiel Hogerheijde for his advice
in setting up and using the RATRAN radiative transfer program,
and Joe McMullin, the referee, for his helpful comments and
suggestions.



\begin{thebibliography}{}
\baselineskip=12pt

\bibitem[Adams(2000)]{adams00}
        {Adams, F.C. 2000, \apj, 542, 964}

\bibitem[Adams \& Fatuzzo(1996)]{adams96}
        {Adams, F.C., \& Fatuzzo, M. 1996, \apj, 464, 256}

\bibitem[Allen(1972)]{allen72}
        {Allen, D. A. 1972, \apj, 172, L55}

\bibitem[Bergin et al.(1997)]{bergin97}
        {Bergin, E.A., Ungerechts, H., Goldsmith, P. F., Snell, R. L.,
         Irvine, W. M., \& Schloerb, F. P. 1997, \apj, 482, 267}

\bibitem[Blaauw(1991)]{blaauw91}
        {Blaauw, A. 1991, in {\it The Physics of Star Formation and
         Early Stellar Evolution}, eds. C.J. Lada \& N.D. Kylafis}

\bibitem[Brown, Dekker, \& de Zeeuw(1997)]{brown97}
        {Brown, A. G. A., Dekker, G., \& de Zeeuw, P. T.
         1997, \mnras, 285, 479}

\bibitem[Hogerheijde \& van der Tak (2000)]{hoger00}
        {Hogerheijde, M. R., \& van der Tak, F. F. S. 2000, A\&A, 362, 697}

\bibitem[Lada, Margulis, \& Dearborn(1984)]{lada84}
        {Lada, C. J., Margulis, M., \& Dearborn, D. 1984, ApJ, 285, 141}

\bibitem[Lada et al.(1993)]{lada93}
        {Lada, C. J., Young, E. T., \& Greene, T. P. 1993, ApJ, 408, 471}

\bibitem[Lada et al.(1994)]{lada94}
        {Lada, C. J., Lada, E. A., Clemens, D. P.,
         \& Bally, J. 1994, ApJ, 429, 694}

\bibitem[Leung \& Brown(1977)]{leung77}
        {Leung, C. M., \& Brown, R. B. 1977, \apj, 214, L73}

\bibitem[Margulis, Lada, \& Snell(1988)]{margulis88}
        {Margulis, M., Lada, C. J., \& Snell, R. L. 1988, \apj, 333, 316}

\bibitem[Margulis, Lada, \& Young(1989)]{margulis89}
        {Margulis, M., Lada, C. J., \& Young, E. T. 1989, \apj, 345, 906}

\bibitem[Menten et al.(1987)]{menten87}
        {Menten, K. M., Serabyn, E., Guesten, R., \& Wilson, T. L.
         1987, \aap, 177, L57}

\bibitem[Mundy, Looney, \& Welch(2000)]{mundy00}
        {Mundy, L.G., Looney, L.W., \& Welch, W.J. 2000,
         in {\it Protostars and Planets IV},
         eds. V. Mannings, A.P. Boss \& S.S. Russell, 355}

\bibitem[Myers et al.(1996)]{myers96}
        {Myers, P. C., Mardones, D., Tafalla, M., Williams, J. P.,
         \& Wilner, D. J. 1996, \apj, 465, L133}

\bibitem[Oliver, Masheder \& Thaddeus(1996)]{oliver96}
        {Oliver, R. J., Masheder, M. R. W.,
         \& Thaddeus, P. 1996, \aap, 315, 578}


\bibitem[Rohlfs \& Wilson(1996)]{rohlfs96}
        {Rohlfs, K., \& Wilson, T. L. 1996, {\it Tools of Radio Astronomy}
         (Heidelburg: Springer)}

\bibitem[Sargent et al.(1984)]{sargent84}
        {Sargent, A. I., van Duinen, R. J., Nordh, H. L., Fridlund, C. V. M.,
         Aalders, J. W., \& Beintema, D. 1984, \aap, 135, 377}

\bibitem[Shu(1977)]{shu77}
        {Shu, F. H. 1977, \apj, 214, 488}

\bibitem[Shu, Adams, \& Liazno(1987)]{shu87}
        {Shu, F. H., Adams, F. C., \& Lizano, S. 1987, \araa, 25, 23}

\bibitem[Sung, Bessell, \& Lee(1997)]{sung97}
        {Sung, H., Bessell, M. S., \& Lee, S. W. 1997, \aj, 114, 2644}

\bibitem[Schreyer et al.(1997)]{schreyer97}
        {Schreyer, K., Helmich, F. P., van Dishoeck, E. F., \& Henning, Th.
         1997, \aap, 326, 347}

\bibitem[Thompson et al.(1998)]{thompson98}
        {Thompson, R. I., Corbin, M. R., Young, E., \& Schneider, G.
         1998, \apj, 492, L177}

\bibitem[Ungerechts et al.(1997)]{ungerechts97}
        {Ungerechts, H., Bergin, E. A., Goldsmith, P. F., Irvine, W. M.,
         Schloerb, F. P., \& Snell, R. L., 1997, \apj, 482, 245}

\bibitem[Velusamy \& Langer(1998)]{velusamy98}
        {Velusamy, T., \& Langer, W. D. 1998, Nature, 392, 685}

\bibitem[Walker et al.(1986)]{walker86}
        {Walker, C. K., Lada, C. J., Young, E. T., Maloney, P. R.,
         \& Wilking, B. A. 1986, \apj, 309, L47}

\bibitem[Ward-Thompson et al.(2000)]{ward00}
        {Ward-Thompson, D., Zylka, R., Mezger, P. G., \& Sievers, A. W.
         2000, \aap, 344, 1122}


\bibitem[Williams \& Myers(1999)]{williams99a}
        {Williams, J. P., \& Myers, P. C. 1999, \apj, 511, 208}

\bibitem[Williams et al.(1999)]{williams99b}
        {Williams, J. P., Myers, P. C., Wilner, D. J., \& Di Francesco, J.
         1999, \apj, 513, L61}

\bibitem[Williams \& Myers(2000)]{williams00}
        {Williams, J. P., \& Myers, P. C. 2000, \apj, 537, 891}

\bibitem[Wolf-Chase \& Gregersen(1997)]{wolf97}
        {Wolf-Chase, G. A., \& Gregersen, E. 1997, \apj, 479, L67}

\bibitem[Wolf-Chase, Walker, \& Lada(1995)]{wolf95a}
        {Wolf-Chase, G. A., Walker, C. K., \& Lada, C. J. 1995, \apj, 442, 197}


\bibitem[Zhou(1995)]{zhou95}
        {Zhou, S. 1995, \apj, 442, 685}

\end{thebibliography}
\end{document}